\documentclass[12pt]{iopart}
\usepackage{graphicx}
\usepackage{latexsym}

\usepackage{iopams}
\def\apj{ApJ\,  }
\def\apjl{ApJ\,  }
\def\apss{Astrophysics and Space Science  }

\def\sn1993j{SN\,1993J~}
\begin{document}
\title
{
Pad\'e  approximant for the equation of motion of 
a supernova remnant
}
\vskip  1cm
\author     {L. Zaninetti}
\address    {
Physics Department,
 via P.Giuria 1,\\ I-10125 Turin,Italy 
}
\ead {zaninetti@ph.unito.it}

\begin {abstract}
In this paper we derive  three equations
of motion for 
a supernova remnant (SNR) in the framework of the
thin layer approximation using 
the Pad\'e  approximant.
The circumstellar medium is
assumed to follow  a density profile of
either an exponential type,
a Gaussian    type,
or  a  Lane--Emden ($n=5$) type.
The three equations of motion  are applied to four 
SNRs: 
Tycho,
Cas A,   
Cygnus loop, 
and  SN~1006.
The percentage error of the Pad\'e  approximated solution 
is always less than $10\%$.
The theoretical decrease of the  velocity over ten years  for  SNRs 
is evaluated.
\end{abstract}
\vspace{2pc}
\maketitle

\section{Introduction}
The equation of motion for a supernova remnant (SNR) 
can be modeled by a single law of motion or multiple
laws of motion when the appropriate boundary 
conditions are provided.
Examples of a single law of motion are:
the Sedov expansion
in 
the presence of a circumstellar medium (CSM) with constant density
where  the radius, $r$,
scales as   $r \propto  t^{0.4}$, see \cite{Sedov1959},
and the momentum conservation in the
framework of the thin layer approximation 
with CSM at constant density    
where   $R \propto  t^{0.25}$,
see \cite{Dyson1997}.
Examples of piece-wise solutions for an SNR can be found 
in \cite{Dalgarno1987}: a first energy conserving phase,
$r \propto  t^{0.4}$ followed by a second adiabatic phase 
where  $r \propto  t^{0.285}$.
At the same time it has been shown 
that in the first ten years of \sn1993j 
$r \propto  t^{0.82}$, which means an observed exponent larger
than the previously suggested exponents, 
see \cite{Zaninetti2011a}.
The previous analysis  allows posing
a basic question: `Is it possible to find an analytical 
solution for SNRs given the three observable astronomical 
parameters, age, radius and velocity ?'.
In order to answer the above question,
Section \ref{secprofiles} introduces three profiles for 
the CSM,
Section \ref{secmotion} derives three 
Pad\'e  approximated laws of motion for SNRs,
and 
Section \ref{secastro}
closes the derived equations of motion  for four SNRs.

\section{Profiles of density}

This  section  introduces three density profiles
for the CSM:
an exponential profile,
a  Gaussian    profile,
and a  self-gravitating profile  of Lane--Emden type.

\label{secprofiles}
\subsection{The exponential profile}

This density  is  assumed to have the following
exponential dependence on $r$
in spherical  coordinates:
\begin{equation}
 \rho(r;r_0,b,\rho_0) =
\rho_0  \exp{(-\frac{(r-r_0)}{b})}
\quad ,
\label{profexponential}
\end{equation}
where $b$ represents the scale.
The piece-wise  density is
\begin{equation}
 \rho (r;r_0,b,\rho_0)  = \left\{ \begin{array}{ll} 
            \rho_0                      & \mbox         {if $r \leq r_0 $ } \\
            \rho_0 \exp{-(\frac{(r-r_0)}{b})}   & \mbox {if $r >    r_0 $ } 
            \end{array}
            \right.
\label{profile_exponential}
\end{equation}
The total mass swept,   $M(r;r_0,b,\rho_0) $,
in the interval $[0,r]$ is
\begin{eqnarray}
M(r;r_0,b,\rho_0) = \nonumber  \\
\frac{4}{3}\,\rho_{{0}}\pi\,{r_{{0}}}^{3}
\nonumber \\
-4\,b \left( 2\,{b}^{2}+2\,br+{r}^{2}
 \right) \rho_{{0}}{{\rm e}^{{\frac {r_{{0}}-r}{b}}}}\pi+4\,b \left( 2
\,{b}^{2}+2\,br_{{0}}+{r_{{0}}}^{2} \right) \rho_{{0}}\pi
\quad .
\end{eqnarray}

\subsection{The Gaussian  profile}

This density  has  the 
Gaussian  dependence 
\begin{equation}
 \rho(r;r_0,b,\rho_0) =
\rho_0  \exp{(-\frac{1}{2}\frac{r^2}{b^2})}
\quad ,
\label{prof_gaussian}
\end{equation}
and the piece-wise  density is
\begin{equation}
 \rho (r;r_0,b,\rho_0)  = \left\{ \begin{array}{ll} 
            \rho_0                      & \mbox         {if $r \leq r_0 $ } \\
            \rho_0\exp{(-\frac{1}{2}\frac{r^2}{b^2})}   & \mbox {if $r >    r_0 $ } 
            \end{array}
            \right.
\end{equation}

The total mass swept,   $M(r;r_0,b,\rho_0) $,
in the interval $[0,r]$ is
\begin{eqnarray}
M(r;r_0,b,\rho_0) = \nonumber  \\
\frac{4}{3}\,\rho_{{0}}\pi\,{r_{{0}}}^{3}+4\,\rho_{{0}}\pi\, \big  ( -{{\rm e}^
{-\frac{1}{2}\,{\frac {{r}^{2}}{{b}^{2}}}}}r{b}^{2}+\frac{1}{2}\,{b}^{3}\sqrt {\pi}
\sqrt {2}{\rm erf}   (\frac{1}{2}\,{\frac {\sqrt {2}r}{b}}  )  \big )\nonumber \\
 -
4\,\rho_{{0}}\pi\, \big  ( -{{\rm e}^{-\frac{1}{2}\,{\frac {{r_{{0}}}^{2}}{{b}^
{2}}}}}r_{{0}}{b}^{2}+\frac{1}{2}\,{b}^{3}\sqrt {\pi}\sqrt {2}{\rm erf}   (
\frac{1}{2}\,{\frac {\sqrt {2}r_{{0}}}{b}}  ) \big  ) 
\quad ,
\end{eqnarray}
where ${\rm erf}$ is the error function,    
see \cite{NIST2010}.

\subsection{The Lane--Emden profile}

The Lane--Emden profile when $n=5$, 
after \cite{Lane1870,Emden1907}, 
is 
\begin{equation}
\rho(r;r_0,b,\rho_0) =\rho_0 {\frac {1}{{(1+ \frac{{r}^{2}}{3b^2})^{\frac{5}{2}}}} }
\label{profile_lane}
\quad ,
\end{equation}
\begin{equation}
 \rho (r;r_0,b,\rho_0)  = \left\{ \begin{array}{ll} 
            \rho_0                      & \mbox         {if $r \leq r_0 $ } \\
            \rho_0 {\frac {1}{{(1+ \frac{{r}^{2}}{3b^2})^{\frac{5}{2}}}} }   & \mbox {if $r >    r_0 $ } 
            \end{array}
            \right.
\end{equation}
The total mass swept,   $M(r;r_0,b,\rho_0) $,
in the interval $[0,r]$ is
\begin{eqnarray}
M(r;r_0,b,\rho_0) = \nonumber  \\
\frac{4}{3}\,\rho_{{0}}\pi\,{r_{{0}}}^{3}+4\,{\frac {{b}^{3}{r}^{3}\rho_{{0}}
\sqrt {3}\pi}{ \left( 3\,{b}^{2}+{r}^{2} \right) ^{\frac{3}{2}}}}-4\,{\frac {{
b}^{3}{r_{{0}}}^{3}\rho_{{0}}\sqrt {3}\pi}{ \left( 3\,{b}^{2}+{r_{{0}}
}^{2} \right) ^{\frac{3}{2}}}}
\quad .
\end{eqnarray}

\section{The equation of motion}
\label{secmotion}
The conservation of the  momentum in
spherical coordinates
in the framework of the thin
layer approximation  states that
\begin{equation}
M_0(r_0) \,v_0 = M(r)\,v
\quad ,
\end{equation}
where $M_0(r_0)$ and $M(r)$ are the masses swept at $r_0$ and $r$,
and $v_0$ and $v$ are the velocities of 
the thin layer at $r_0$ and $r$.

\subsection{Motion with exponential profile}

Assuming an exponential profile as given by 
Eq.~(\ref{profile_exponential})
the velocity is
\begin{equation}
\frac{dr}{dt} = \frac{NE}{DE}
\quad ,
\label{vel_exp}
\end{equation}
where
\begin{eqnarray}
NE =
{-{r_{{0}}}^{3}v_{{0}}}
\nonumber  
\quad,
\end{eqnarray}
and  
\begin{eqnarray}
DE=
6\,{{\rm e}^{{\frac {r_{{0}}-r}{b}}}}{b}^{3}+6\,{{\rm e}^{{\frac {r_{{0
}}-r}{b}}}}{b}^{2}r
\nonumber \\
+3\,{{\rm e}^{{\frac {r_{{0}}-r}{b}}}}b{r}^{2}-{r_{
{0}}}^{3}-3\,{r_{{0}}}^{2}b-6\,r_{{0}}{b}^{2}-6\,{b}^{3}
\quad .
\nonumber
\end{eqnarray}
In the above differential equation of the first order in $r$, the
variables can be separated and integration
gives the following non-linear equation:
\begin{eqnarray}
\frac {1}{{r_{{0}}}^{3}{\it v_0}}
\bigg (
18\,{{\rm e}^{{\frac {r_{{0}}-r}{b}}}}{b}^{4}+12\,{{\rm e}^{{\frac {r_
{{0}}-r}{b}}}}{b}^{3}r+3\,{{\rm e}^{{\frac {r_{{0}}-r}{b}}}}{b}^{2}{r}
^{2}-{r_{{0}}}^{4}-3\,{r_{{0}}}^{3}b
\nonumber\\
+{r_{{0}}}^{3}r-9\,{r_{{0}}}^{2}{b
}^{2}+3\,{r_{{0}}}^{2}br-18\,{b}^{3}r_{{0}}+6\,r_{{0}}{b}^{2}r-18\,{b}
^{4}+6\,{b}^{3}r
\bigg )
\nonumber \\
=\left( t-{\it t_0} \right) 
\label .
\label{eqn_nl_exp}
\end{eqnarray}
In this case is not possible to find an analytical  
solution for the radius, $r$,
as a function of  time.
We therefore apply 
the Pad\'e rational polynomial 
approximation of degree 2 in the numerator
and degree 1 in the denominator about the point $r=r_0$ 
to the  left-hand  side of 
Eq.~(\ref{eqn_nl_exp}):
\begin{equation}
\frac
{
- \left( r_{{0}}-r \right)  \left( -5\,br-br_{{0}}-2\,rr_{{0}}+2\,{r_{
{0}}}^{2} \right) 
}
{
2\,v_{{0}} \left( 2\,br-5\,br_{{0}}-rr_{{0}}+{r_{{0}}}^{2} \right)  
}
= t-t_0 
\quad .
\end{equation}
The resulting Pad\'e  approximant  for the radius $r_{2,1}$ is
\begin{eqnarray}
r_{2,1}=\frac{1}{2\,r_{{0}}+5\,b}
\Bigg  ( r_{{0}}tv_{{0}}-r_{{0}}{\it t_0}\,v_{{0}}-2\,btv_{{0}}+2\,b{\it t_0}\,v_
{{0}}+2\,{r_{{0}}}^{2}+2\,r_{{0}}b
\nonumber \\
+ \biggl ( {4\,{b}^{2}{t}^{2}{v_{{0}}}^{
2}-8\,{b}^{2}t{\it t_0}\,{v_{{0}}}^{2}+4\,{b}^{2}{{\it t_0}}^{2}{v_{{0}}
}^{2}-4\,b{t}^{2}r_{{0}}{v_{{0}}}^{2}}
\nonumber\\
{
+8\,bt{\it t_0}\,r_{{0}}{v_{{0}}}^
{2}-4\,b{{\it t_0}}^{2}r_{{0}}{v_{{0}}}^{2}+{t}^{2}{r_{{0}}}^{2}{v_{{0}
}}^{2}-2\,t{\it t_0}\,{r_{{0}}}^{2}{v_{{0}}}^{2}+{{\it t_0}}^{2}{r_{{0}}
}^{2}{v_{{0}}}^{2}
}
\nonumber \\
{
+42\,{b}^{2}tr_{{0}}v_{{0}}
-42\,{b}^{2}{\it t_0}\,r_{
{0}}v_{{0}}+6\,bt{r_{{0}}}^{2}v_{{0}}-6\,b{\it t_0}\,{r_{{0}}}^{2}v_{{0
}}+9\,{r_{{0}}}^{2}{b}^{2}} \biggl )^{\frac{1}{2}} 
\Bigg )
\quad ,
\label{rmotionexp} 
\end{eqnarray}
and the velocity is
\begin{equation}
v_{2,1}=\frac{dr_{2,1}}{dt} =\frac{NVE}{DVE}
\quad ,
\label{vmotionexp} 
\end{equation}
\begin{eqnarray} 
NVE =
4\,v_{{0}} \Big  \{    ( -b/2+1/4\,r_{{0}}   )\times  \nonumber \\
\sqrt {4\,   ( 
b-\frac{1}{2}\,r_{{0}}   ) ^{2}   ( t-t_{{0}}   ) ^{2}{v_{{0}}}^{2}
+42\,   ( b+1/7\,r_{{0}}   )    ( t-t_{{0}}   ) br_{{0}}
v_{{0}}+9\,{r_{{0}}}^{2}{b}^{2}}+\nonumber  \\  ( 3/4\,b+   ( t/4-1/4\,t_{{0
}}   ) v_{{0}}   ) {r_{{0}}}^{2}+{\frac {21\,r_{{0}}b}{4}
   ( v_{{0}}   ( -{\frac {4\,t}{21}}+{\frac {4\,t_{{0}}}{21}}
   ) +b   ) }
\nonumber  \\
+{b}^{2}v_{{0}}   ( t-t_{{0}}   ) 
  \Big \}
\quad ,
\end{eqnarray}
and
\begin{eqnarray}
DVE =   \nonumber \\
\sqrt {4\,   ( b-\frac{1}{2}\,r_{{0}}   ) ^{2}   ( t-t_{{0}}
   ) ^{2}{v_{{0}}}^{2}+42\,   ( b+1/7\,r_{{0}}   )    ( 
t-t_{{0}}   ) br_{{0}}v_{{0}}+9\,{r_{{0}}}^{2}{b}^{2}} \times
\nonumber \\
  ( 2\,r
_{{0}}+5\,b   ) 
\quad .
\end{eqnarray}

\subsection{Motion with Gaussian profile}

Assuming a   Gaussian  profile as given by 
Eq.~(\ref{prof_gaussian}) 
the velocity is
\begin{equation}
\frac{dr}{dt} = \frac{NG}{DG}
\quad ,
\label{vel_gaussian}
\end{equation}
where
\begin{eqnarray}
NG= -2\,{r_{{0}}}^{3}v_{{0}}
\end{eqnarray}
and  
\begin{eqnarray}
DG= 
-3\,{b}^{3}\sqrt {\pi}\sqrt {2}{\rm erf} \left(\frac{1}{2}\,{\frac {\sqrt {2}r
}{b}}\right)
\nonumber \\
+3\,{b}^{3}\sqrt {\pi}\sqrt {2}{\rm erf} \left(\frac{1}{2}\,{
\frac {\sqrt {2}r_{{0}}}{b}}\right)+6\,{{\rm e}^{-\frac{1}{2}\,{\frac {{r}^{2}
}{{b}^{2}}}}}r{b}^{2}
\nonumber \\
-6\,{{\rm e}^{-\frac{1}{2}\,{\frac {{r_{{0}}}^{2}}{{b}^{2
}}}}}r_{{0}}{b}^{2}-2\,{r_{{0}}}^{3}
\quad .
\end{eqnarray}
The appropriate  non-linear equation  is
\begin{eqnarray}
\frac{1}{2\,{r_{{0}}}^{3}v_{{0}}} 
\bigg (
  ( -12\,{b}^{4}+6\,r_{{0}}   ( r-r_{{0}}   ) {b}^{2}
   ) {{\rm e}^{-\frac{1}{2}\,{\frac {{r_{{0}}}^{2}}{{b}^{2}}}}}+12\,{b}^{4
}{{\rm e}^{-\frac{1}{2}\,{\frac {{r}^{2}}{{b}^{2}}}}}
\nonumber \\
-3\,\sqrt {\pi}{\rm erf} 
  (\frac{1}{2}\,{\frac {\sqrt {2}r_{{0}}}{b}}  )\sqrt {2}{b}^{3}r+3\,{b
}^{3}\sqrt {\pi}\sqrt {2}{\rm erf}   (\frac{1}{2}\,{\frac {\sqrt {2}r}{b}}
  )r
\nonumber \\
+2\,{r_{{0}}}^{3}   ( r-r_{{0}}   )
\bigg )
= t-t_0 \, . 
\label{eqn_nl_gaussian}
\end{eqnarray}
The Pad\'e rational polynomial 
approximation of degree 2 in the numerator
and degree 1 in the denominator 
about $r=r_0$ 
for the left-hand side of the above equation gives
\begin{eqnarray}
\frac
{
1
}
{
2\,v_{{0}} \left( 2\,{b}^{2}r-5\,r_{{0}}{b}^{2}-r{r_{{0}}}^{2}+{r_{{0}
}}^{3} \right)
}
\Bigg (
-   ( r-r_{{0}}   )   \bigg ( 9\,{{\rm e}^{-\frac{1}{2}\,{\frac {{r_{{0}}
}^{2}}{{b}^{2}}}}}{b}^{2}r
\nonumber \\
-9\,{{\rm e}^{-\frac{1}{2}\,{\frac {{r_{{0}}}^{2}}{{
b}^{2}}}}}r_{{0}}{b}^{2}-4\,{b}^{2}r+10\,r_{{0}}{b}^{2}+2\,r{r_{{0}}}^
{2}-2\,{r_{{0}}}^{3} \bigg  )
\Bigg )
= t-t_0 \, .
\end{eqnarray}
The resulting Pad\'e  approximant  for the radius $r_{2,1}$ is
\begin{eqnarray}
r_{2,1}=  
\frac{1}
{
9\,{{\rm e}^{-\frac{1}{2}\,{\frac {{r_{{0}}}^{2}}{{b}^{2}}}}}{b}^{2}+2\,{r_{{0
}}}^{2}-4\,{b}^{2}
}
\Bigg \{
9\,{{\rm e}^{-\frac{1}{2}\,{\frac {{r_{{0}}}^{2}}{{b}^{2}}}}}r_{{0}}{b}^{2}-2
\,{b}^{2}tv_{{0}}
\nonumber \\
+2\,{b}^{2}{\it t_0}\,v_{{0}}+{r_{{0}}}^{2}tv_{{0}}-{r
_{{0}}}^{2}{\it t_0}\,v_{{0}}-7\,r_{{0}}{b}^{2}+2\,{r_{{0}}}^{3}
\nonumber \\
+ \bigg [ 
{54\,{b}^{4}r_{{0}}v_{{0}} \Big( t-{\it t_0} \Big) {{\rm e}^{-\frac{1}{2}\,{
\frac {{r_{{0}}}^{2}}{{b}^{2}}}}}
}
\nonumber \\
{
+4\, \Big(  \Big( t-{\it t_0}
 \Big)  \Big( {b}^{2}-\frac{1}{2}\,{r_{{0}}}^{2} \Big) v_{{0}}-\frac{3}{2}\,r_{{0
}}{b}^{2} \Big) ^{2}}
\bigg ]^{\frac{1}{2}}
\Bigg \}  
\label{rmotiongauss}
\, , 
\end{eqnarray}
and the velocity is
\begin{equation}
v_{2,1}=\frac{dr_{2,1}}{dt} =\frac{NVG}{DVG}
\quad ,
\label{vmotiongauss} 
\end{equation}
\begin{eqnarray} 
NVG =
-   \Bigg ( -27\,{{\rm e}^{-\frac{1}{2}\,{\frac {{r_{{0}}}^{2}}{{b}^{2}}}}}r_{{0}
}{b}^{4}+   ( 2\,{b}^{2}-{r_{{0}}}^{2}   )    ( v_{{0}}
   ( t-t_{{0}}   ) {r_{{0}}}^{2}
\nonumber \\
+3\,r_{{0}}{b}^{2}
-2\,v_{{0}}{b
}^{2}   ( t-t_{{0}}   ) 
+ \bigg \{ {54\,{b}^{4}r_{{0}}v_{{0}}
   ( t-t_{{0}}   ) {{\rm e}^{-\frac{1}{2}\,{\frac {{r_{{0}}}^{2}}{{b}^{
2}}}}}}
\nonumber \\
{ 
+4\,   (    ( t-t_{{0}}   )    ( {b}^{2}-\frac{1}{2}\,{r_{{0
}}}^{2}   ) v_{{0}}-3/2\,r_{{0}}{b}^{2}   ) ^{2}}   ) 
  \bigg \} ^{\frac{1}{2}} \Bigg ) v_{{0}}
\quad ,
\end{eqnarray}
and
\begin{eqnarray} 
DVG =
\Bigg \{ {54\,{b}^{4}r_{{0}}v_{{0}}   ( t-t_{{0}}   ) {{\rm e}^{-1
/2\,{\frac {{r_{{0}}}^{2}}{{b}^{2}}}}}+4\,   (    ( t-t_{{0}}
   )    ( {b}^{2}-\frac{1}{2}\,{r_{{0}}}^{2}   ) v_{{0}}
}
\nonumber \\
{
-3/2\,r_{{0
}}{b}^{2}   ) ^{2}}   
\Bigg \}^{\frac{1}{2}}
( 9\,{{\rm e}^{-\frac{1}{2}\,{\frac {{r_{{0}}}^{2
}}{{b}^{2}}}}}{b}^{2}+2\,{r_{{0}}}^{2}-4\,{b}^{2}   ) 
\quad .
\end{eqnarray}

\subsection{Motion with the Lane--Emden profile}

Assuming a Lane--Emden profile, $n=5$,     as given by 
Eq.~(\ref {profile_lane}),
the velocity is
\begin{equation}
\frac{dr}{dt} = \frac{NL}{DL}
\quad ,
\label{vel_lane}
\end{equation}
where
\begin{eqnarray}
NL = {r_{{0}}}^{3}v_{{0}} \left( 3\,{b}^{2}+{r}^{2} \right) ^{\frac{3}{2}} \left( 3
\,{b}^{2}+{r_{{0}}}^{2} \right) ^{\frac{3}{2}}
\end{eqnarray}
and 
\begin{eqnarray}
DL = -3\, \left( 3\,{b}^{2}+{r}^{2} \right) ^{\frac{3}{2}}\sqrt {3}{r_{{0}}}^{3}{b}
^{3}+3\, \left( 3\,{b}^{2}+{r_{{0}}}^{2} \right) ^{\frac{3}{2}}\sqrt {3}{b}^{3
}{r}^{3}
\nonumber \\
+ \left( 3\,{b}^{2}+{r}^{2} \right) ^{\frac{3}{2}} \left( 3\,{b}^{2}+{
r_{{0}}}^{2} \right) ^{\frac{3}{2}}{r_{{0}}}^{3}
\, . 
\end{eqnarray}
The connected  non-linear equation  is
\begin{eqnarray}
\frac {1}
{
{r_{{0}}}^{3}v_{{0}} \left( 3\,{b}^{2}+{r_{{0}}}^{2} \right) ^{\frac{3}{2}}
\sqrt {3\,{b}^{2}+{r}^{2}}
}
 \times 
\nonumber \\
\bigg (
54\,   ( {b}^{2}+\frac{1}{3}\,{r_{{0}}}^{2}   )    ( \frac{1}{18}\,{r_{{0}}}
^{3}   ( r-r_{{0}}   ) \sqrt {3\,{b}^{2}+{r}^{2}}
\nonumber \\
+{b}^{3}\sqrt 
{3}   ( {b}^{2}+\frac{1}{6}\,{r}^{2}   )    ) \sqrt {3\,{b}^{2}+{r_
{{0}}}^{2}}-54\,\sqrt {3\,{b}^{2}+{r}^{2}}\sqrt {3}{b}^{3}   ( {b}^
{4}
\nonumber \\ 
+\frac{1}{2}\,{b}^{2}{r_{{0}}}^{2}+\frac{1}{18}\,r{r_{{0}}}^{3}   )
\bigg ) =t-t_0
\nonumber  
\quad .
\end{eqnarray}
The Pad\'e rational polynomial 
approximation of degree 2 in the numerator
and degree 1 in the denominator for the left-hand side of the above equation gives
\begin{eqnarray}
\frac
{
NP
}
{
2\, \left( 3\,{b}^{2}+{r_{{0}}}^{2} \right) ^{\frac{3}{2}}v_{{0}} \left( 2\,r{
b}^{2}-5\,{b}^{2}r_{{0}}-r{r_{{0}}}^{2} \right) 
}
=t-t_0
\, ,
\end{eqnarray}
where 
\begin{eqnarray}
PN =
-27\,   ( r-r_{{0}}   )  
  \Big ( \big  ( -\frac{4}{9}\,   ( r{b}^{2}-\frac{5}{2}\,{b}
^{2}r_{{0}}-\frac{1}{2}\,r{r_{{0}}}^{2} \big  ) \times
\nonumber \\
  \big ( {b}^{2}+\frac{1}{3}\,{r_{{0}}}
^{2}  \big ) \sqrt {3\,{b}^{2}+{r_{{0}}}^{2}}+{b}^{5}\sqrt {3}  \big ( 
r-r_{{0}} \big  )   \Big  ) 
\, .
\end{eqnarray}
The  Pad\'e  approximant  for the radius is 
\begin{eqnarray}
r_{2,1}=\frac{NR}{DR}
\label{rmotionlaneemden}   
\end{eqnarray}
where 
\begin{eqnarray}
NR= -18\,   ( {b}^{2}+\frac{1}{3}\,{r_{{0}}}^{2}    ) ^{2}{b}^{2} 
  ( -\frac{1}{2}\,{r_{{0}}}^{3}-\frac{1}{2}\,v_{{0}}   ( t-t_{{0}}    ) {r_{{0}}}^{2}
\nonumber \\
+ \frac{7}{2} \,{b}^{2}r_{{0}}+{b}^{2}v_{{0}}   ( t-t_{{0}}    )     ) 
\sqrt {3\,{b}^{2}+{r_{{0}}}^{2}}+   ( 81\,{b}^{9}r_{{0}}+27\,{b}^{7
}{r_{{0}}}^{3}    ) \sqrt {3}
\nonumber \\
+\sqrt {972}
\Bigg ( {   ( {b}^{2}+\frac{1}{3}
\,{r_{{0}}}^{2}    ) ^{4}{b}^{4}   ( \frac{9}{2}\,\sqrt {3}r_{{0}}{b}^{5
}v_{{0}}   ( t-t_{{0}}    ) \sqrt {3\,{b}^{2}
+{r_{{0}}}^{2}}
}
\nonumber \\
{
+
 \bigg  ( -\frac{1}{2}\,{r_{{0}}}^{3}-\frac{1}{2}\,v_{{0}}   ( t-t_{{0}}    ) {r_{
{0}}}^{2}-\frac{3}{2}\,{b}^{2}r_{{0}}
}
\nonumber \\
{
+{b}^{2}v_{{0}}   ( t-t_{{0}}    ) 
    ) ^{2}   ( {b}^{2}+\frac{1}{3}\,{r_{{0}}}^{2}    )  \bigg   ) } \Bigg )^{\frac{1}{2}}
\, ,
\end{eqnarray}
and  
\begin{eqnarray}
DR=
{b}^{2}   ( 3\,{b}^{2}+{r_{{0}}}^{2}   ) \bigg   ( 27\,{b}^{5}
\sqrt {3}-12\,{b}^{4}\sqrt {3\,{b}^{2}+{r_{{0}}}^{2}}
\nonumber \\
+2\,{b}^{2}{r_{{0
}}}^{2}\sqrt {3\,{b}^{2}+{r_{{0}}}^{2}}+2\,{r_{{0}}}^{4}\sqrt {3\,{b}^
{2}+{r_{{0}}}^{2}} \bigg  ) 
\quad ,
\end{eqnarray}
and the velocity is
\begin{equation}
v_{2,1}=\frac{dr_{2,1}}{dt} =\frac{NVL}{DVL}
\quad ,
\label{vmotionlaneemden}
\end{equation}
where 
\begin{eqnarray} 
NVL =
-18\,\sqrt {3}   ( 3\,{b}^{2}+{r_{{0}}}^{2}   ) v_{{0}} 
  \bigg  (  
  \bigg  ( -243\,   ( {b}^{2}+\frac{1}{3}\,{r_{{0}}}^{2}   ) ^{2}{b}^{7}r_
{{0}}\sqrt {3}
\nonumber \\
+\sqrt {972}
{ \Bigg \{  ( {b}^{2}+\frac{1}{3}\,{r_{{0}}}^{2}
   ) ^{4}{b}^{4}   ( 9/2\,\sqrt {3}r_{{0}}{b}^{5}v_{{0}}
   ( t-t_{{0}}   ) \sqrt {3\,{b}^{2}+{r_{{0}}}^{2}}
}
\nonumber \\
{
+   ( {b}
^{2}+\frac{1}{3}\,{r_{{0}}}^{2}   ) \bigg   ( -\frac{1}{2}\,{r_{{0}}}^{3}-\frac{1}{2}\,v_{{0
}}   ( t-t_{{0}}   ) {r_{{0}}}^{2}-3/2\,{b}^{2}r_{{0}}
}
\nonumber  \\
{
+{b}^{2}v
_{{0}}   ( t-t_{{0}}   )    ) ^{2} \bigg  ) } 
\Bigg \}^{\frac{1}{2}}
  ( 2\,{b}^
{2}-{r_{{0}}}^{2}   ) \Bigg   ) \sqrt {3\,{b}^{2}+{r_{{0}}}^{2}}
\nonumber  \\
-
108\,   ( {b}^{2}+\frac{1}{3}\,{r_{{0}}}^{2}   ) ^{3}{b}^{2}   ( -1/
2\,{r_{{0}}}^{3}-\frac{1}{2}\,v_{{0}}   ( t-t_{{0}}   ) {r_{{0}}}^{2}-3
/2\,{b}^{2}r_{{0}}
\nonumber  \\
+{b}^{2}v_{{0}}   ( t-t_{{0}}   )    ) 
   ( {b}^{2}-\frac{1}{2}\,{r_{{0}}}^{2} ) 
\quad ,
\end{eqnarray}
and
\begin{eqnarray} 
DVL =
18\,\sqrt {972}\sqrt {3}
\Bigg 
\{ 
  ( {b}^{2}+\frac{1}{3}\,{r_{{0}}}^{2}
   ) ^{4}{b}^{4}   ( 9/2\,\sqrt {3}r_{{0}}{b}^{5}v_{{0}}
\nonumber \\
   ( t-t_{{0}}   ) \sqrt {3\,{b}^{2}+{r_{{0}}}^{2}}+   ( {b}
^{2}+\frac{1}{3}\,{r_{{0}}}^{2}   )    ( -\frac{1}{2}\,{r_{{0}}}^{3}-\frac{1}{2}\,v_{{0
}}   ( t-t_{{0}}   ) {r_{{0}}}^{2}
\nonumber  \\
-3/2\,{b}^{2}r_{{0}}+{b}^{2}v
_{{0}}   ( t-t_{{0}}   )    ) ^{2}   ) 
\Bigg 
\}
^{\frac{1}{2}} 
  ( 
   ( -12\,{b}^{4}+2\,{b}^{2}{r_{{0}}}^{2}+2\,{r_{{0}}}^{4}   ) 
\sqrt {3\,{b}^{2}+{r_{{0}}}^{2}}
\nonumber \\
+27\,{b}^{5}\sqrt {3}   ) 
\quad .
\end{eqnarray}

\section{Astrophysical Applications}
\label{secastro}
In the previous section, we derived three equations of motion 
in the form of non-linear  equations 
and three Pad\'e approximated equations of motion.
We now check the reliability of the numerical and approximated 
solutions on four SNRs: Tycho, see \cite{Williams2016},
Cas A, see \cite{Patnaude2009},  Cygnus loop,  see \cite{Chiad2015},
and  SN~1006, see \cite{Uchida2013}.
The three astronomical measurable parameters 
are the time since the explosion in years, $t$,
the actual observed radius in pc, $r$,
and the present velocity of expansion in 
km\,s$^{-1}$, see Table \ref{tablesnrs}.
\begin{table}[ht!]
\caption {
Observed astronomical parameters of SNRs
}
\label{tablesnrs}
\begin{center}
\begin{tabular}{|c|c|c|c|c|}
\hline
Name  & Age (yr)   &  Radius (pc) & Velocity (km\,s$^{-1}$)& References\\
\hline                                                   
Tycho        & 442        &  3.7         & 5300 & Williams~et~al.~2016    \\
Cas ~A       & 328        &  2.5         & 4700 & Patnaude~and~Fesen~2009 \\
Cygnus~loop  & 17000      &  24.25       & 250  & Chiad~et~al.~2015       \\
SN ~1006     & 1000       &  10.19       & 3100 & Uchida~et~al.2013       \\
\hline
\end{tabular}
\end{center}
\end{table}
The astrophysical  units have not yet been specified:
pc for length  and  yr for time
are the units most commonly used by astronomers.
With these units, the initial velocity is 
$v_0(km s^{-1})= 9.7968 \, 10^5 v_0(pc\,yr^{-1})$.
The determination of the four  unknown parameters, which are   
$t_0$, $r_0$, $v_0$ and $b$,  
can be obtained by equating the observed astronomical velocities 
and radius with those obtained with the 
Pad\'e rational polynomial, i.e. 
\begin{eqnarray}
\label{eqnnl1}
        r_{2,1}&= Radius(pc),\\
\label{eqnnl2}
        v_{2,1}&=Velocity(km s^{-1}). 
\end{eqnarray}

In order to reduce the unknown parameters from four to two, we
fix $v_0$ and  $t_0$.
The two parameters $b$ and $r_0$ are found by solving the
two non-linear equations (\ref{eqnnl1})  
and (\ref{eqnnl2}).
The results for the three types of profiles here adopted 
are reported in Tables 
\ref{tablesnrsexp},
\ref{tablesnrsgauss}
and
\ref{tablesnrslaneemden}.

\begin{table}[ht!]
\caption {
Theoretical  parameters of SNRs
for the Pad\'e  approximated equation of motion 
with an exponential profile. 
}
\label{tablesnrsexp}
\begin{center}
\begin{tabular}{|c|c|c|c|c|c|c|}
\hline
Name         &$t_0$(yr)&$r_0$(pc)&$v_0(km\,s^{-1})$& b(pc)&$\delta\,(\%)$&
$\Delta\,v (km\,s^{-1}) $ \\
\hline                                                   
Tycho        &  0.1  &  1.203    & 8000   &  0.113  &   5.893  & -1.35   \\
Cas ~A       &  1    &  0.819    & 8000   &  0.1    &   6.668  & -3.29   \\
Cygnus~loop  &  10   &  12.27    & 3000   & 45.79   &   6.12   & -0.155  \\
SN ~1006     &  1    &  5.49     & 3100   & 2.332   &   1.455  & -12.34  \\
\hline
\end{tabular}
\end{center}
\end{table}

\begin{table}[ht!]
\caption {
Theoretical  parameters of SNRs
for the Pad\'e  approximated equation of motion 
with a Gaussian  profile. 
}
\label{tablesnrsgauss}
\begin{center}
\begin{tabular}{|c|c|c|c|c|c|c|}
\hline
Name         &$t_0$(yr)&$r_0$(pc)&$v_0(km\,s^{-1})$& b(pc)&$\delta\,(\%)$&
$\Delta\,v (km\,s^{-1}) $ \\
\hline                                                   
Tycho        &  0.1  &  1.022    & 8000   &  0.561   &   8.517  & -10.469   \\
Cas ~A       &  1    &  0.741    & 7000   &  0.406   &   7.571  & -13.16   \\
Cygnus~loop  &  10   &  11.92    & 3000   &  21.803  &   7.875  & -0.161  \\
SN ~1006     &  1    &  5.049    & 10000  &  4.311   &   4.568  & -18.58  \\
\hline
\end{tabular}
\end{center}
\end{table}

\begin{table}[ht!]
\caption 
{
Theoretical  parameters of SNRs
for the Pad\'e  approximated equation of motion 
with a  Lane--Emden profile. 
}
\label{tablesnrslaneemden}
\begin{center}
\begin{tabular}{|c|c|c|c|c|c|c|}
\hline
Name         &$t_0$(yr)&$r_0$(pc)&$v_0(km\,s^{-1})$& b(pc)&$\delta\,(\%)$&
$\Delta\,v (km\,s^{-1}) $ \\
\hline                                                   
Tycho        &  0.1  &   0.971    & 8000   & 0.502   &    3.27   & -14.83   \\
Cas ~A       &  1    &   0.635    & 8000   & 0.35    &    4.769  & -23.454   \\
Cygnus~loop  &  10   &   11.91    & 3000   & 27.203  &    7.731  & -0.162  \\
SN ~1006     &  1    &   5        & 10000  & 4.85    &    3.297  & -19.334   \\
\hline
\end{tabular}
\end{center}
\end{table}
The    goodness of the approximation is evaluated
through the percentage error, $\delta$, which is
\begin{equation}
\delta = \frac{\big | r_{2,1} - r_E \big |}
{r_E} \times 100
\quad ,
\end{equation}
where $r_{2,1}$ is the 
Pad\'e approximated radius and
$r_E$ is the exact solution which is obtained by solving numerically
the non-linear equation of motion, as an example Eq.~(\ref{eqn_nl_exp})
in the exponential case.
The numerical values of $\delta$ are reported 
in  column 6 of  
Tables 
\ref{tablesnrsexp},
\ref{tablesnrsgauss}
and
\ref{tablesnrslaneemden}.
Another useful astrophysical variable  is the predicted decrease in velocity 
on the basis of the     
Pad\'e approximated velocity, $v_{2,1}$, in 10 years,
see column 7 of  
Tables 
\ref{tablesnrsexp},
\ref{tablesnrsgauss}
and
\ref{tablesnrslaneemden}.

\section{Conclusions}

The expansion of an SNR can be modeled by the conservation
of momentum in the presence of a decreasing density:
here we analysed an exponential, a Gaussian 
and a Lane--Emden profile. 
The three equations of motion have complicated left-hand sides 
but simple left-hand sides, viz.,  $(t-t_0)$.
The application of the 
Pad\'e  approximant to the left-hand side of the complicated equation of motion 
allows finding three approximate laws of motion,
see Eqs~(\ref{rmotionexp}, \ref{rmotiongauss}, \ref{rmotionlaneemden}),
and three approximate velocities, 
see Eqs~(\ref{vmotionexp}, \ref{vmotiongauss}, \ref{vmotionlaneemden}). 
The astrophysical test is performed 
on four spherical SNRs assumed to be spherical
and the four sets  of  parameters  are 
reported in  Tables  
\ref{tablesnrsexp},
\ref{tablesnrsgauss}
and
\ref{tablesnrslaneemden}.
The percentage of error of the Pad\'e  approximated 
solutions for the radius is always less than 
$10\%$ with respect to the numerical exact solution,
see column 6 of  
the three last tables.
In order to produce an astrophysical prediction,
the theoretical decrease in velocity for the four SNRs here
analysed is evaluated,
see column 7 of  
Tables 
\ref{tablesnrsexp},
\ref{tablesnrsgauss}
and
\ref{tablesnrslaneemden}.

\noindent
{\bf REFERENCES}

\providecommand{\newblock}{}

\providecommand{\newblock}{}
\end{document}